\documentclass[]{tPHL2e}

\begin{document}
\doi{10.1080/0950083YYxxxxxxxx}
 \issn{1362-3036}
\issnp{0950-0839} 
\jvol{00} \jnum{00} \jyear{2010} \jmonth{Spring}

\markboth{Taylor \& Francis and I.T. Consultant}{Philosophical Magazine Letters}

\articletype{manuscript}

\title{Isothermal remanent magnetization and the spin dimensionality of spin glasses}

\author{Roland Mathieu$^{\rm a}$$^{\ast}$\thanks{$^\ast$Corresponding author. Email: roland.mathieu@angstrom.uu.se}, Matthias Hudl$^{\rm a}$, Per Nordblad$^{\rm a}$, Yusuke Tokunaga$^{\rm b}$,\\
Yoshio Kaneko$^{\rm b}$, Yoshinori Tokura$^{\rm b,c}$, Hiroko Aruga Katori$^{\rm d}$, Atsuko Ito$^{\rm e}$\\
\vspace{6pt} $^{\rm a}${\em{Department of Engineering Sciences, Uppsala University, Box 534, SE-751 21 Uppsala, Sweden}};\\
\vspace{0pt}$^{\rm b }${\em{Multiferroics Project, ERATO-JST, Tokyo 113-8656, Japan}};\\
\vspace{0pt}$^{\rm c}${\em{Department of Applied Physics, University of Tokyo, Tokyo 113-8656, Japan}};\\
\vspace{0pt}$^{\rm d}${\em{Magnetic Materials Laboratory, RIKEN, 2-1 Hirosawa, Wako, Saitama 351-0198, Japan}};\\
\vspace{0pt}$^{\rm e}${\em{Advanced Meson Science Laboratory, RIKEN, 2-1 Hirosawa, Wako, Saitama 351-0198, Japan}}.\\
\vspace{6pt}\received{2010} }

\maketitle

\begin{abstract}

The isothermal remanent magnetization is used to investigate dynamical magnetic properties of spatially three dimensional spin glasses with different spin dimensionality ($Ising$, $XY$, $Heisenberg$). The isothermal remanent magnetization is recorded vs. temperature after intermittent application of a weak magnetic field at a constant temperature $T_h$. We observe that in the case of the $Heisenberg$ spin glasses, the equilibrated spin structure and the direction of the excess moment are recovered at $T_h$. The isothermal remanent magnetization thus reflects the directional character of the Dzyaloshinsky-Moriya interaction present in $Heisenberg$ systems.

\bigskip

\end{abstract}

\section{Introduction}

The macroscopic response of the spin-glass phase to a magnetic field change is linear at low enough fields \cite{orm}. Experiments on the time and temperature dependence of the spin-glass magnetization can be used to learn about the underlying spin configuration and its spontaneous re-organization towards more favorable states (aging) \cite{ghost}.  The response function, $R(t,ta)$ reflects the spin re-organization and appears stationary at observation times log($t$) $<$ log ($t_a$) and non-stationary when  log($t$) $\sim$ log ($t_a$), where $t_a$ is the effective age of the spin glass and $t$ the observation time of the experimental probe. The apparent age of the system is governed by the time spent at constant temperature (wait time, $t_w$) or in a temperature dependent experiment by the cooling/heating rate.\\
In this study we have considered three model spin glasses of spatial dimensionality three, but with different spin dimensionality 1, 2 and 3: Fe$_{0.5}$Mn$_{0.5}$TiO$_3$ ($Ising$) \cite{Ising}, Eu$_{0.5}$Sr$_{1.5}$MnO$_4$ ($XY$) \cite{xy} and Cu(Mn) ($Heisenberg$) \cite{cumn}. For all systems, we have studied the evolution and memory of the spin configuration that has been imprinted by a halt at constant temperature, $T_h$, during cooling. The probe we have used in this study is the isothermal remanent magnetization (IRM) that is attained by intermittently applying a magnetic field at the halt temperature. In the $Heisenberg$ systems, a local and random Dzyaloshinsky-Moriya (DM) interaction\cite{DM} is inherently at play \cite{fert}. For comparison, we have also investigated a Au(Fe) system, another three dimensional $Heisenberg$ spin glass, however, with stronger apparent anisotropy \cite{dorothee,eric}.

\section{Experimental}

\begin{figure}
\begin{center}
\subfigure[]{
\resizebox*{10cm}{!}{\includegraphics{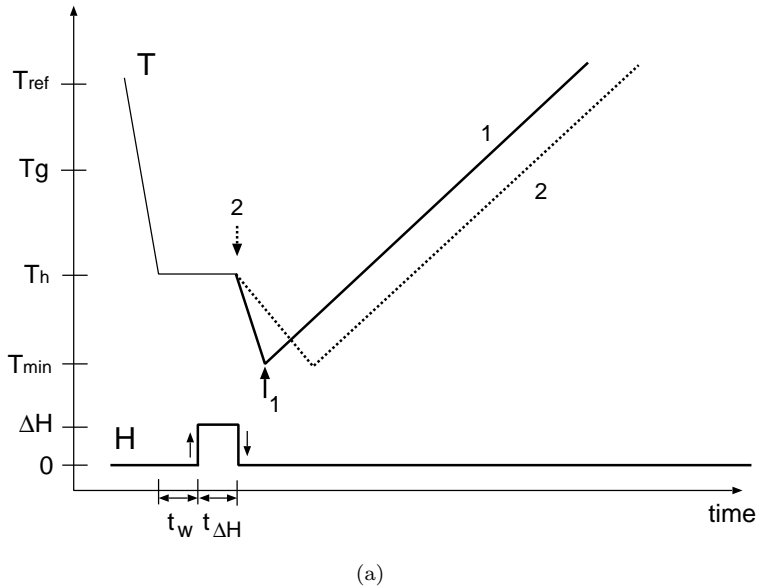}}}%
\caption{Schematic representation of the thermal and magnetic field protocols used to record the Isothermal remanent magnetization (IRM) curves: Variation of the temperature $T$ and magnetic field $H$ as a function of time according to protocol (1) (continuous line) and (2) (dotted line). The arrows indicate the time a which the magnetization starts to be recorded in each case. The initial cooling from a reference temperature $T_{ref}$ to the lowest temperature $T_{min}$ is halted at $T_h$ below the spin-glass transition temperature $T_g$ for a wait time $t_w$ (3 s or 3000 s). A magnetic field of $\Delta H$=1 mT is then applied for $t_{\Delta H}$=3000 s.}%
\label{fig-sketch}
\end{center}
\end{figure}

The probe materials are those listed in the introduction: an $Ising$ Fe$_{0.5}$Mn$_{0.5}$TiO$_3$ single crystal, $T_g$ $\approx$ 21 K), an $XY$ (single crystal, $T_g$ $\approx$ 18 K) and two $Heisenberg$ (Cu(Mn) polycrystalline, $T_g$ $\approx$ 57 K and Au(Fe) polycrystalline, $T_g$ $\approx$ 24 K ) spin glasses, all of spatial dimensionality three \cite{Ising,xy,cumn}.\\
The experiments were performed in an MPMS Superconducting Quantum Interference Device (SQUID) magnetometer with the low magnetic field option, allowing a weak, $<$ 0.05 mT, residual field.\\ 
The temperature dependence of the zero-field-cooled (ZFC), field-cooled (FC) and isothermal remanent magnetization (IRM) of the samples was recorded using a probe field of 1 mT. The isothermal remanent magnetization (IRM) was acquired by intermittently applying a 1 mT field during $t_{\Delta H}$=3000 s at constant temperature $T_h$, ($T_h/T_g$ $\approx$ 0.7) after two different wait times $t_w$ (3 and 3000 s), and measured vs. temperature in two different thermal protocols  illustrated in Fig.~\ref{fig-sketch}. In the first procedure (1), the sample is immediately cooled (10 K/min) to a lowest temperature and the IRM is measured on heating (1 K/min). In the second procedure (2), the sample is cooled immediately after the IRM has been acquired and measured on both cooling (1 K/min) and heating (1 K/min).\\
The magnetic field employed in these experiments (1 mT) is weak enough to yield linear response; this implies that the global evolution (aging) of the spin configuration at constant temperature evolves ``undisturbed'' by this weak perturbation and that the system recovers a closely similar spin state at $T_h$, after the cooling protocol (memory).   

\section{Results}

\begin{figure}
\begin{center}
\subfigure[]{
\resizebox*{6cm}{!}{\includegraphics{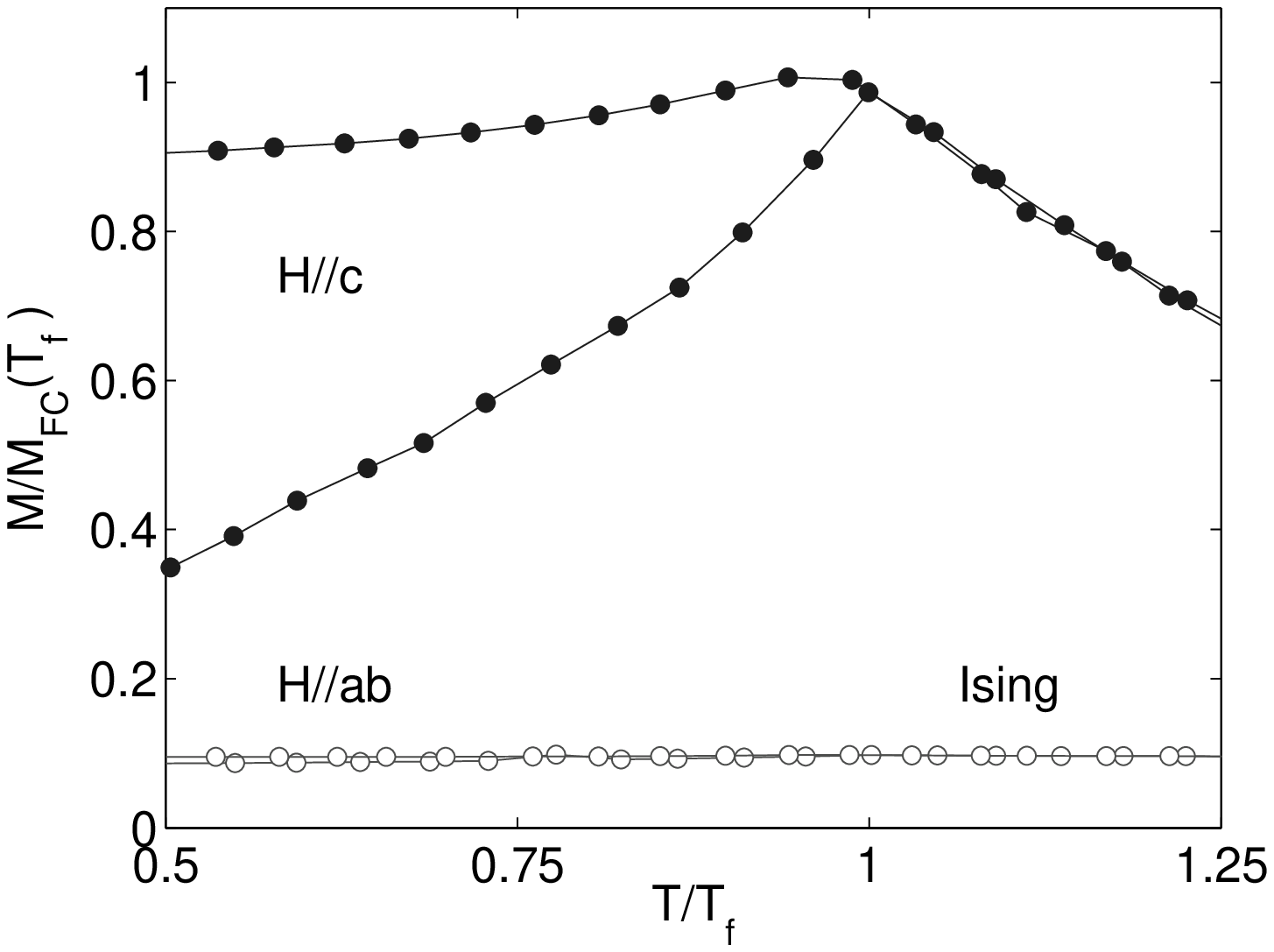}}}%
\subfigure[]{
\resizebox*{6cm}{!}{\includegraphics{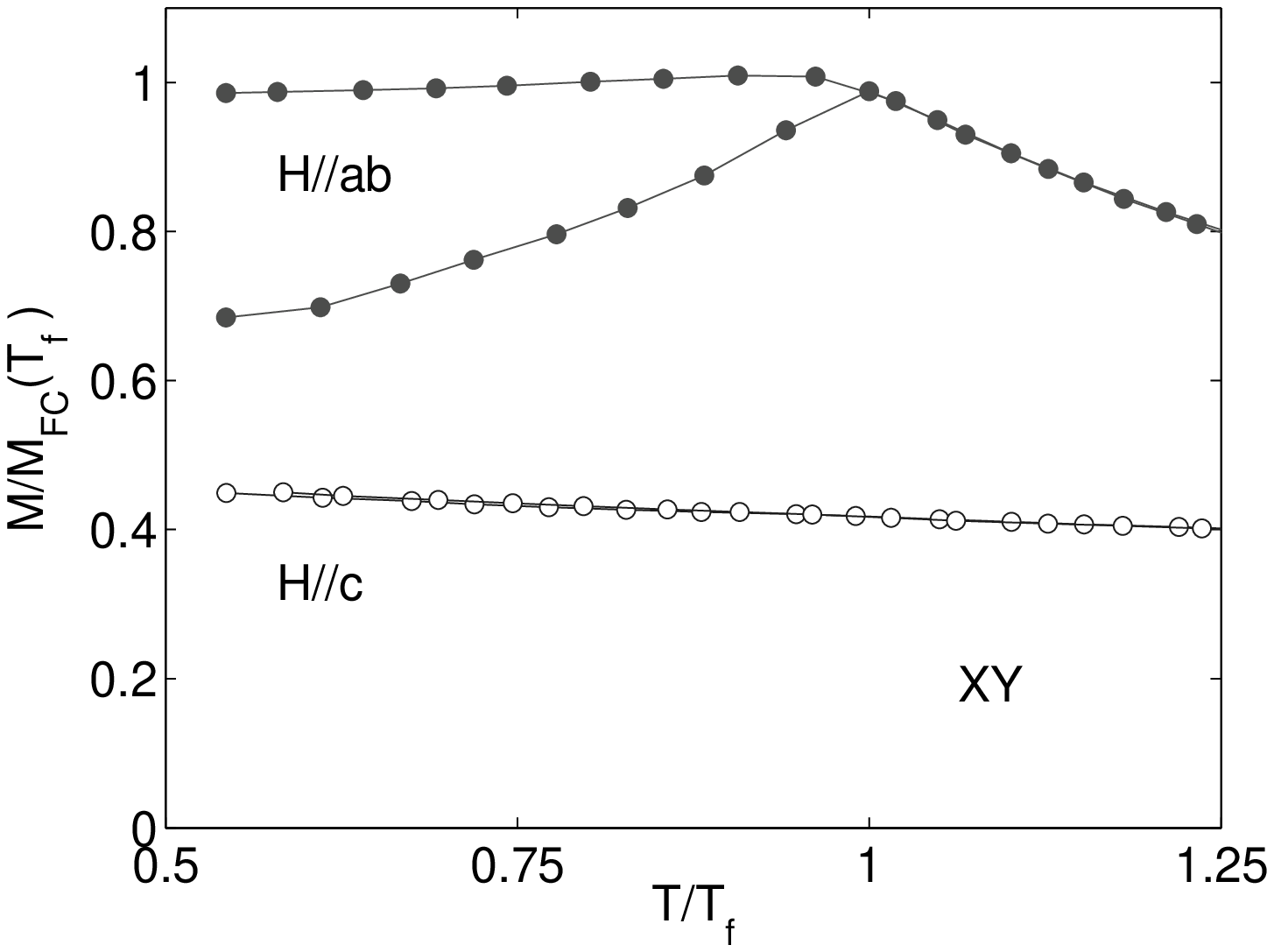}}}\\%
\subfigure[]{
\resizebox*{6cm}{!}{\includegraphics{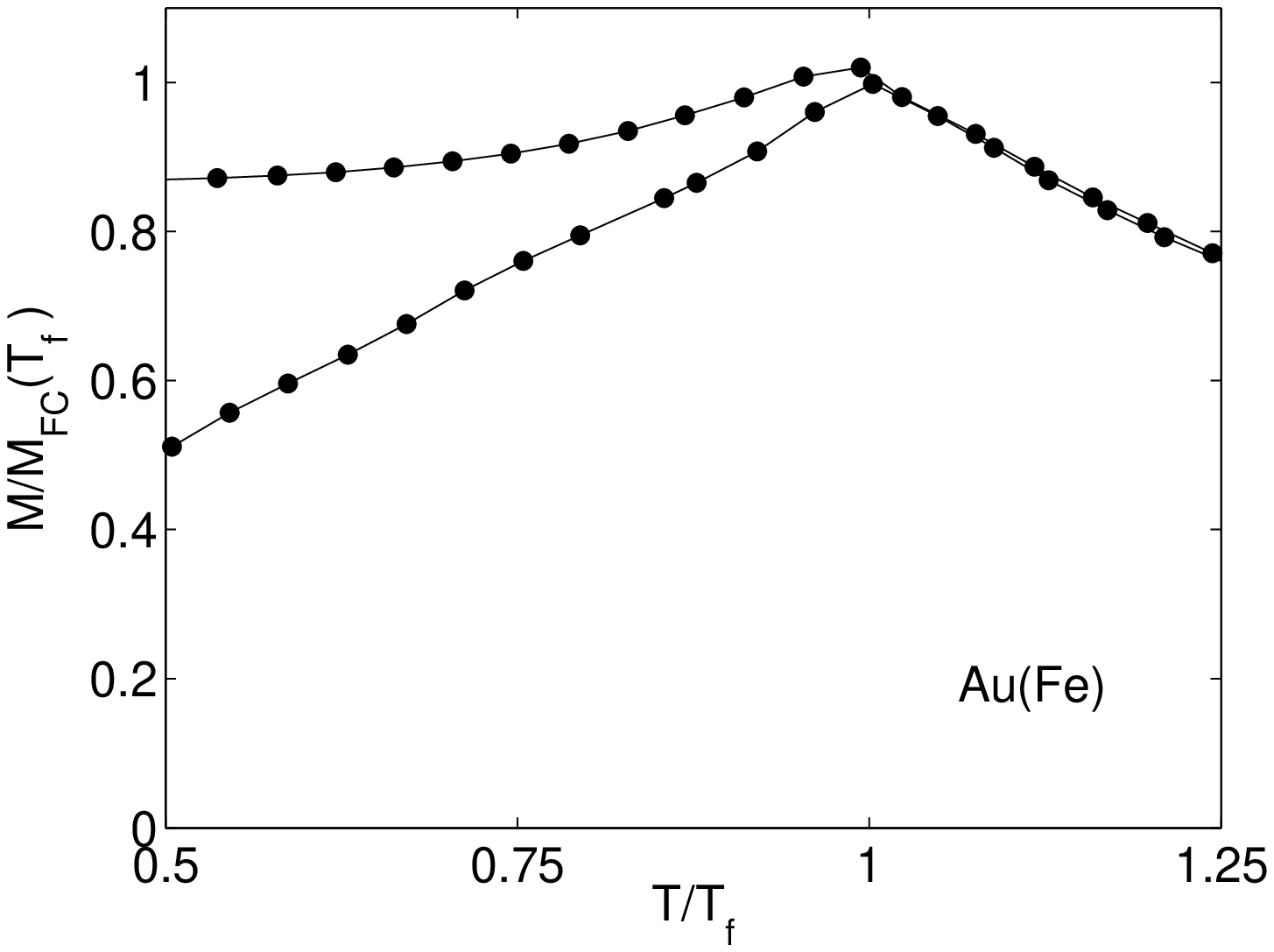}}}%
\subfigure[]{
\resizebox*{6cm}{!}{\includegraphics{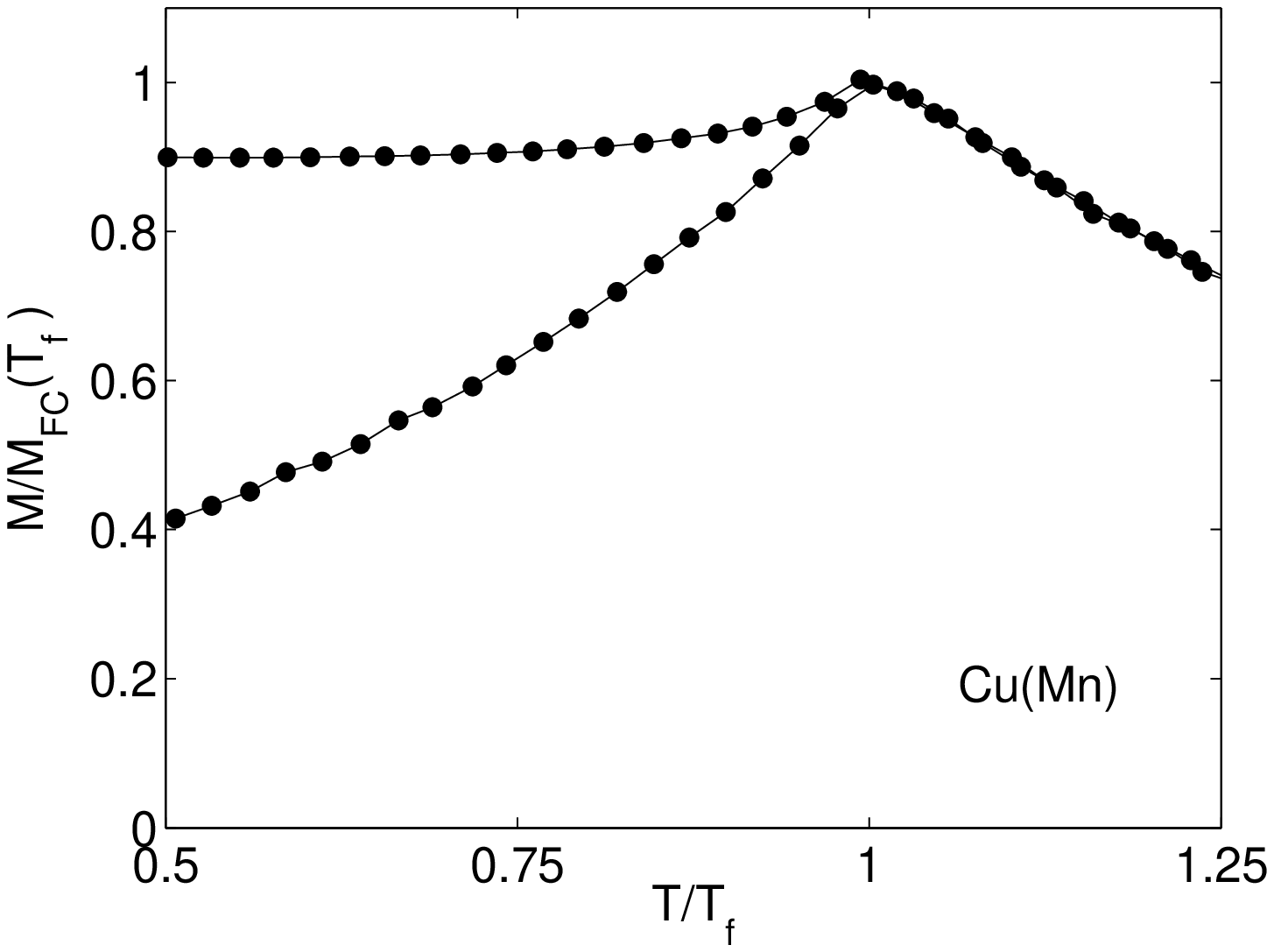}}}\\%
\caption{Temperature $T$ dependence of the zero-field cooled (ZFC) and field-cooled (FC) magnetization $M$ recorded for all systems using a small magnetic field $H$ = 1 mT. $T$ is normalized by the freezing temperature $T_f$ $\approx$ $T_g$, while $M$ is normalized by the field-cooled magnetization value at $T_f$, $M_{FC}(T_f)$.}%
\label{fig-zfcfc}
\end{center}
\end{figure}

\begin{figure}
\begin{center}
\subfigure[]{
\resizebox*{6cm}{!}{\includegraphics{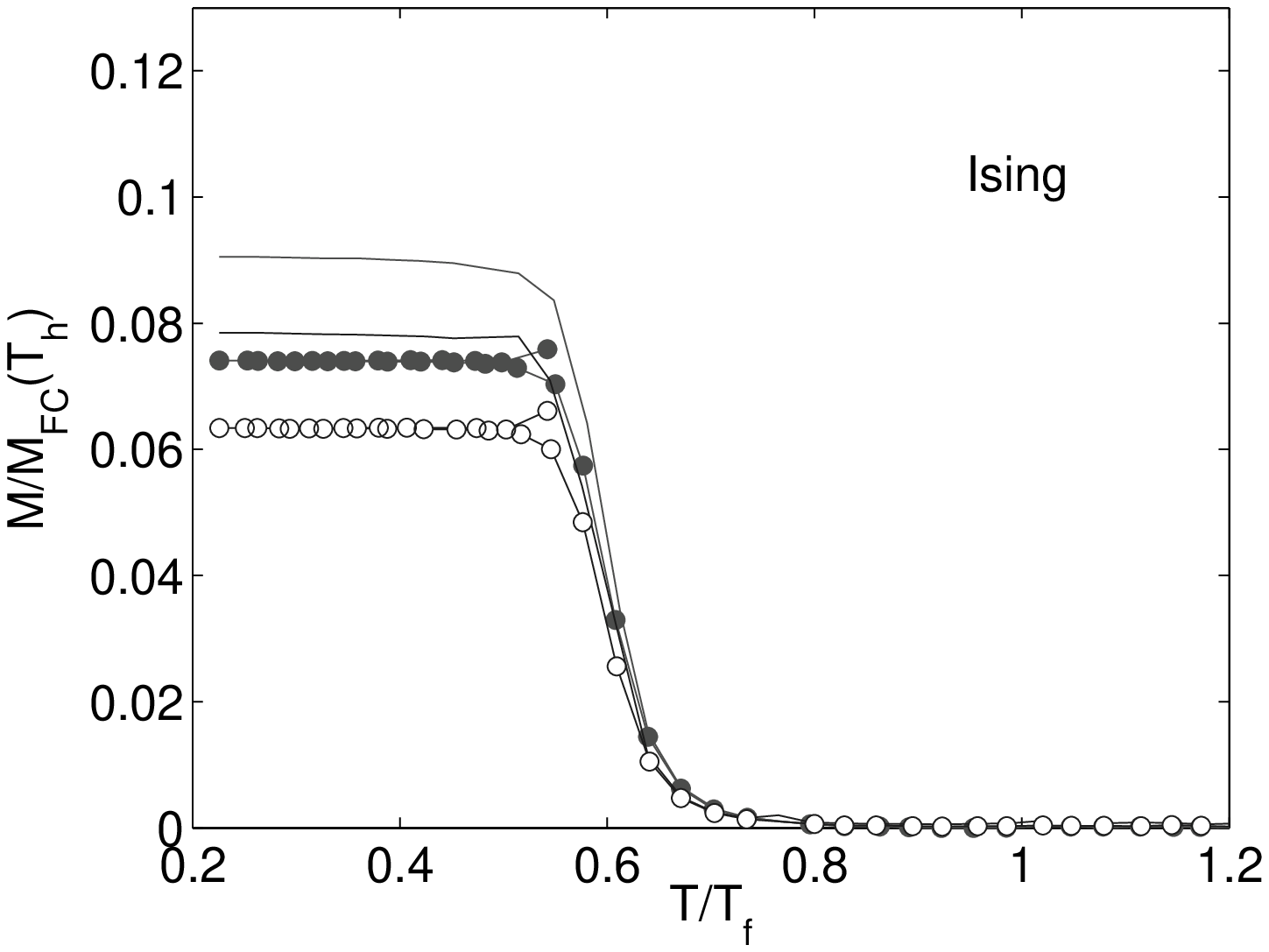}}}%
\subfigure[]{
\resizebox*{6cm}{!}{\includegraphics{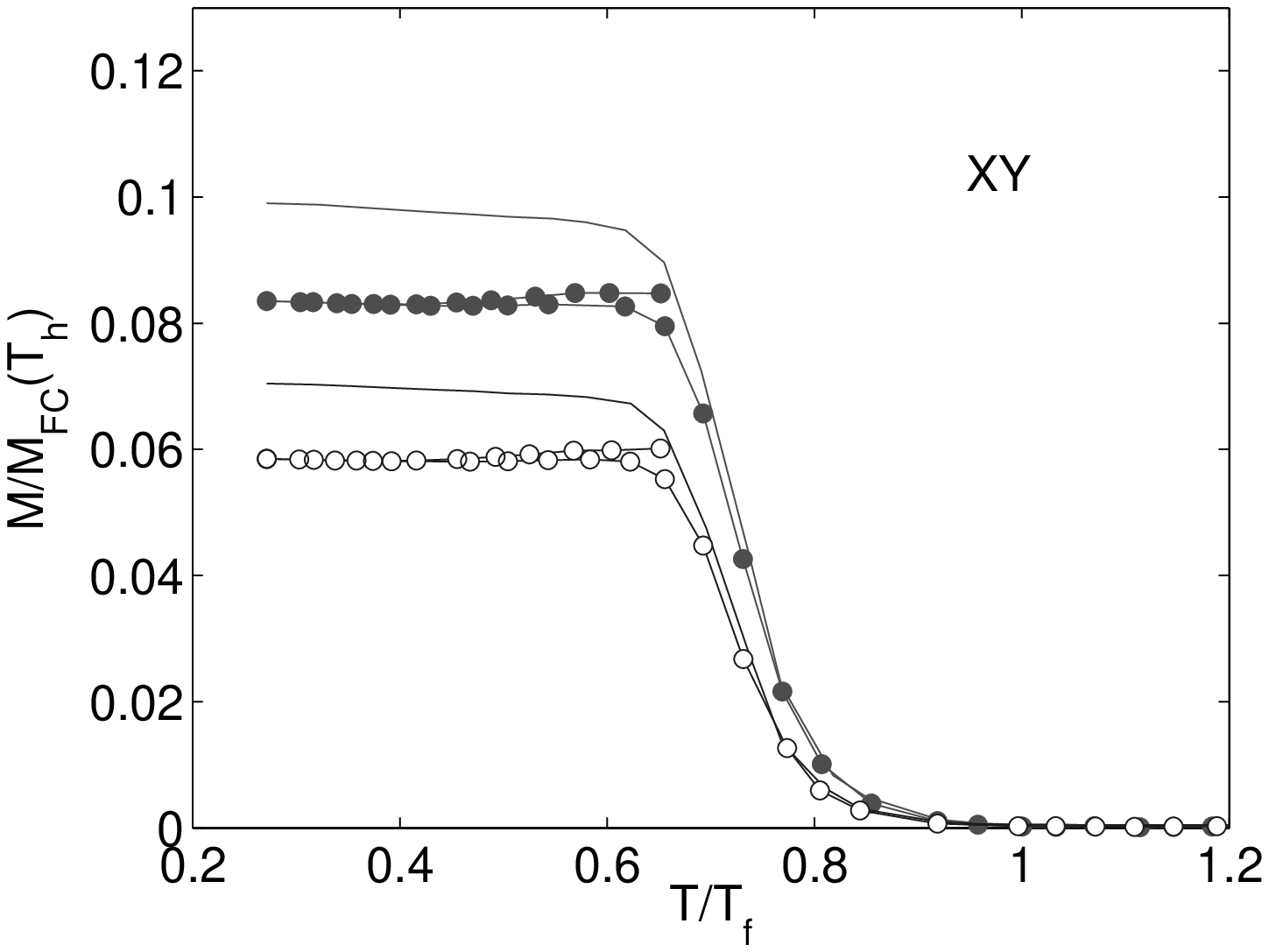}}}\\%
\subfigure[]{
\resizebox*{6cm}{!}{\includegraphics{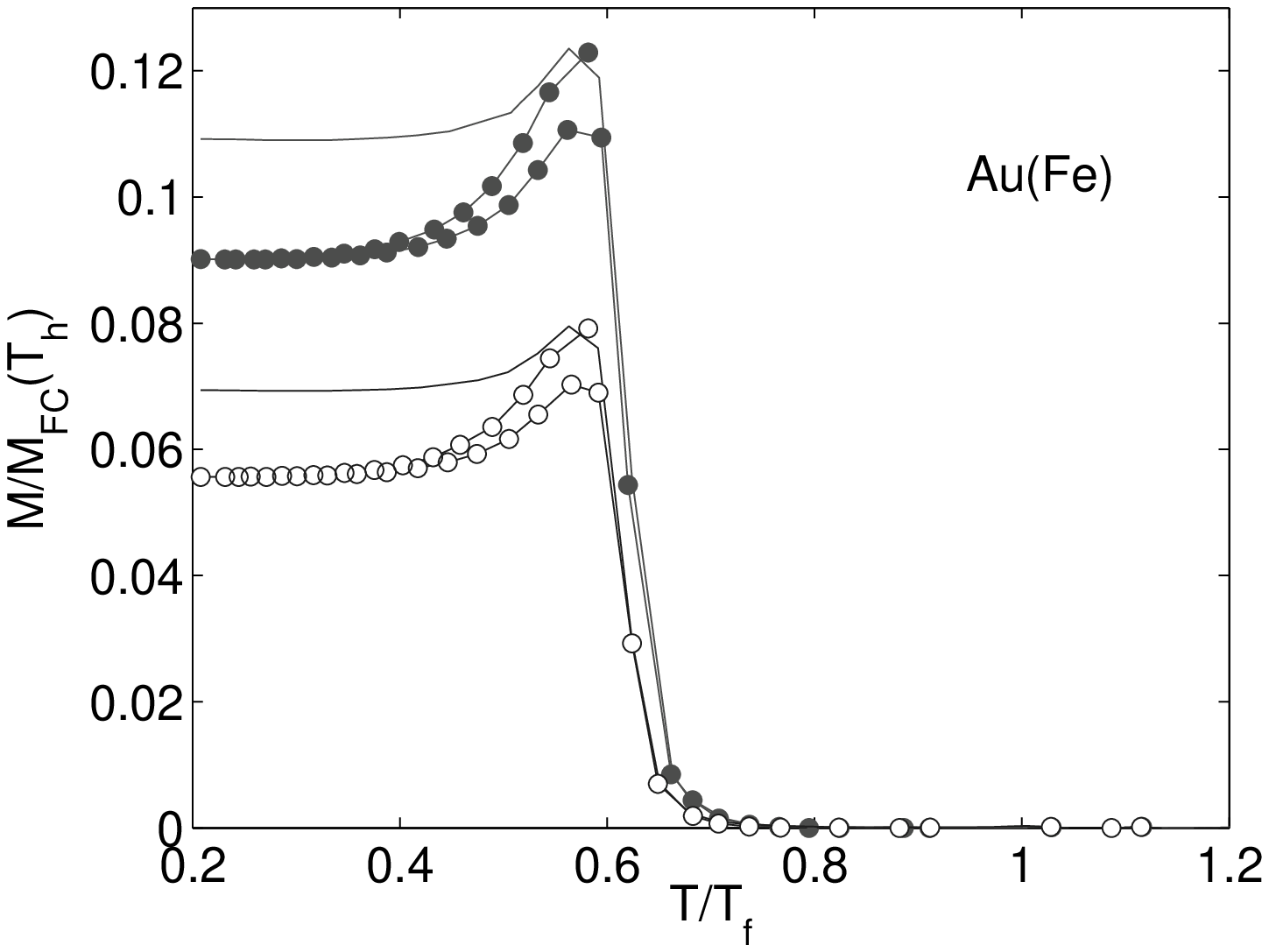}}}%
\subfigure[]{
\resizebox*{6cm}{!}{\includegraphics{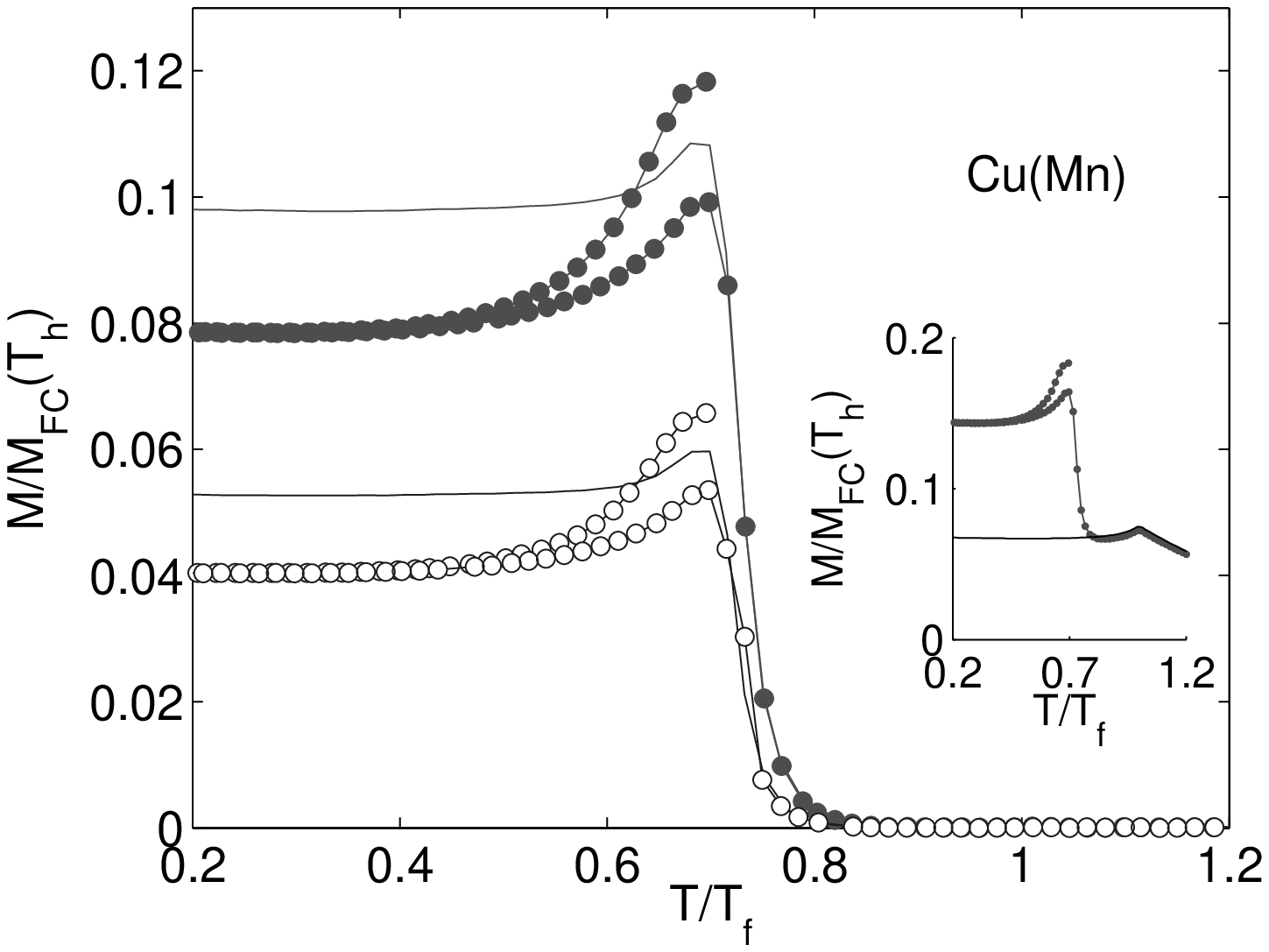}}}\\%
\caption{Temperature dependence of the Isothermal remanent magnetization (IRM), recorded employing the two procedures schematically illustrated in Fig.~\ref{fig-sketch}. $T$ is normalized by the freezing temperature $T_f$, while $M$ is normalized by the value of the field-cooled magnetization in 1mT at the halt temperature $T_h$, $M_{FC}(T_h)$, extracted from Fig.~\ref{fig-zfcfc}. Magnetization data collected using procedure (1) are plotted in continuous lines while the data collected using procedure (2) is plotted using filled ($t_w$=3 s) and open ($t_w$=3000 s) symbols. In all panels, the upper continuous line corresponds to experiments performed with $t_w$=3 s, while the lower one depicts the results of experiments with $t_w$=3000 s. $\Delta H$ is 1mT and $t_{\Delta H}$ is 3000 s in all experiments. In the measurements on Cu(Mn), the reference temperature $T_{ref}$ is 70 K, the halt temperature $T_{h}$ is 40 K, and the lowest measurement temperature $T_{min}$ is 10 K. For Au(Fe), $T_{ref}$=40 K, $T_{h}$=14 K, and $T_{min}$=5 K, while for the $Ising$ and $XY$ systems, $T_{ref}$=30 K, $T_{h}$=12 K, and $T_{min}$=5 K. The inset in panel (d) shows the ``raw'' magnetization data collected following procedure (2), with wait time of 3 s (same filled symbols as in the main frame), as well as a reference background measurement, in which the system is cooled and measured in zero applied magnetic field (continuous line).}%
\label{fig-irm}
\end{center}
\end{figure}

Figure~\ref{fig-zfcfc} shows the temperature dependence of the ZFC, FC and IRM magnetization of the four systems in reduced units. The $Ising$ sample is measured along the direction of the spins (the $c$-axis of the Ilmenite structure) and perpendicular to this axis, in which case the susceptibility is much lower, reversible and only weakly temperature dependent (c.f. Fig.~\ref{fig-zfcfc}(a)). The $XY$ sample is measured in the $ab$-plane of the layered tetragonal structure and perpendicular to this plane, in which case, the susceptibility is much lower, reversible and only weakly temperature dependent (c.f. Fig.~\ref{fig-zfcfc}(b)). In the following we consider and perform measurements on the $Ising$ and $XY$ systems only in the spin directions, i.e. along the $c$-axis for the $Ising$ system, and perpendicular to it for the $XY$ system. The two $Heisenberg$ samples have isotropic magnetic properties; the samples have cylindrical shape and were measured along the cylinder axis, and perpendicular to it for comparison. Similar results were obtained in the two cases; only the data collected along the cylinder axes are shown in Fig.~\ref{fig-zfcfc}(c-d). The overall behavior of the susceptibility curves of the four samples show the characteristic spin-glass features: a cusp in the ZFC curve, irreversibility between the FC and ZFC curves below the cusp, and a weakly temperature dependent FC curve below the irreversibility onset.\\
Figure~\ref{fig-irm} illustrates the temperature dependence of the isothermal remanent magnetization for the different samples. A superposed ``FC'' magnetization associated with the weak residual field in the magnet has been subtracted from the data in all cases; the magnitude of this is of the order of the IRM, see inset of Fig.~\ref{fig-irm}(d), where the measured ``raw'' data of the magnetization of the Cu(Mn) sample (IRM and background) is plotted. The fact that the nominally zero field in the magnetometer in reality is a weak finite field does not affect the IRM results of Fig.~\ref{fig-irm}; since the principle of superposition is valid for the recorded magnetisation response to a field change in the linear response regime \cite{orm}.\\ 
There are common features in the IRM for all four samples. The IRM is frozen in on cooling and shows only a weak temperature dependence on cooling and heating at low temperatures, and fades away rapidly as the temperature is increased above the temperature where it was acquired. A rather similar behavior of the IRM would be observed also for a system of non-interacting (or interacting) magnetic nanoparticles with a distribution of particle sizes \cite{sasaki}. Also in common is a wait time dependence of the magnitude of the IRM; a larger magnitude is obtained for 3 s wait time than for 3000 s wait time at $T_h$ before the 1 mT field is applied for 3000 s. This difference is explained by the wait time dependence of the response function; the increase of magnetization gets slower with wait time at constant temperature in spin glasses (aging) \cite{aging}. Also in common, is that independently of the wait time, the IRM measured on heating after rapid cooling has larger amplitude than after slower cooling.\\ 
There are however significant differences between the behavior of the IRM for the different samples, possibly emanating from the spin dimensionality. The wait time at $T_h$ (3 s and 3000 s) influences the magnitude of the IRM in the sequence $Heisenberg$ - $XY$ - $Ising$, going from strongest to weakest influence. This is in accord with the observation that the influence of aging on the response function follows the same sequence \cite{ghost,xy}. The cooling heating curves for the $XY$ and $Ising$ systems coalesce more rapidly below the halt temperature than the corresponding curves for the $Heisenberg$ systems, irrespective of wait time at $T_h$. 
The most striking difference is the pronounced maximum in the IRM when returning to $T_h$ in Cu(Mn) and Au(Fe). This feature does not appear in the other two samples, where the IRM smoothly decays with temperature through and above $T_h$. Also, significantly, the heating curve in the protocol (2) where the sample is measured both on cooling and heating approaches closer to the curve measured only on heating (protocol (1)) in the neighborhood of $T_h$. This behavior of the $Heisenberg$ system implies that there is an additional memory component in the spin structure that is imprinted during the halt at $T_h$ \cite{epl} (see also \cite{dcagmn}). 

\section{Discussion and Conclusions}

The spin structure of a spin glass is continuously re-organized at low temperature to adjust toward a structure that is most favorable at each temperature. This re-organization occurs always, and irrespective of whether a weak field is applied or a weak field is applied or removed. This implies that the magnetization that is recorded in an experiment may be composed of several independent superposed contributions. The pure IRM (as recorded during procedure (1)) mirrors the excess magnetization that is acquired from an intermittent field application at constant temperature. This magnetization is frozen in on cooling to lower temperatures. During the time it takes to cool to low temperature and again heat the sample back towards $T_h$, the spin structure is spontaneously reorganized on short time-length scales, however, a memory of the structure attained during the stop at $T_h$ remains \cite{ghost,dcagmn}.  During this process also parts of the IRM are lost due to relaxation and this is reflected in a lower value of the magnetization when $T_h$ is recovered. For the $XY$ and $Ising$ systems this picture accords with the observations in Fig.~\ref{fig-irm}, and one needs to make an ordinary memory experiment to realize that the spin configuration at $T_h$ does carry a memory of the equilibration that occurred during the original cooling \cite{ghost,xy}.\\ 
The $Heisenberg$ systems on the other hand, show an additional directional memory of the equilibrated spin structure imprinted at $T_h$. Although a part of the IRM apparently (Fig.~\ref{fig-irm}(a) and (b)) is lost due to the relaxation and restructuring that occurs during cooling and heating (the relaxation is larger for Cu(Mn) than Au(Fe)), a large part of the original IRM magnetization is recovered when $T_h$ is approached as the sample is reheated. This pronounced increase of the IRM is surprising since it is not observed in the other two samples, where the part of the magnetization that has decayed during cooling heating remains lost also at $T_h$.  In the $Heisenberg$ systems on the other hand both the equilibrated spin structure and the direction of the excess moment are recovered at $T_h$.  This property of the Cu(Mn) and Au(Fe) spin glass possibly reflects a reminiscence of the directional character of the Dzyaloshinsky-Moriya interaction inherent to the $Heisenberg$ systems \cite{fert}.\\
Interestingly, it was shown earlier that the critical exponents associated with the $XY$ spin-glass phase transition had values between those of the $Heisenberg$ and $Ising$ systems \cite{xy}, and that ac or dc memory experiments suggested dynamical properties closer to those of $Ising$ systems. It is interesting to note that here the dynamical behavior of the $XY$ system probed by the IRM experiments is qualitatively similar to that of the $Ising$ system.

\section{Conclusion}

We have investigated the dynamical magnetic properties of spatially three dimensional spin glasses by performing measurements of the isothermal remanent magnetization (IRM). Qualitative differences in the IRM curves of the different systems, which have different spin dimensionality (1: $Ising$, 2: $XY$, or 3: $Heisenberg$), were observed. Our results suggest that in $Heisenberg$ systems, the equilibrated spin structure and the direction of the excess moment of the IRM are kept in memory, reflecting the directional character of the (random and local) Dzyaloshinsky-Moriya interaction inherently present in $Heisenberg$ systems.

\subsection*{Acknowledgements}

We thank the Swedish Research Council (VR) and the G\"oran Gustafsson Foundation for financial support.

\newpage


\label{lastpage}

\end{document}